\documentclass[preprint,prd,nofootinbib,showpacs,rotate]{revtex4}

\usepackage{graphicx}
\usepackage{dcolumn}
\usepackage{bm}
\usepackage{color}
\input{epsf}
\begin{document}

\title{Heavy quark potential at finite temperature from gauge/string duality  }

\author{Henrique Boschi-Filho}
\email{boschi@if.ufrj.br}
\affiliation{Instituto de F\'{\i}sica, 
Universidade Federal do Rio de Janeiro, Caixa Postal 68528, 
RJ 21941-972 -- Brazil}
\author{Nelson R. F. Braga}
\email{braga@if.ufrj.br}
\affiliation{Instituto de F\'{\i}sica, 
Universidade Federal do Rio de Janeiro, Caixa Postal 68528, 
RJ 21941-972 -- Brazil}
\author{Cristine N. Ferreira}
\email{crisnfer@if.ufrj.br}
\affiliation{Instituto de F\'{\i}sica,~Universidade Federal do Rio de 
Janeiro, Caixa Postal 68528, 
RJ 21941-972 - Brazil}
\affiliation{ N\'ucleo de F\'isica, Centro Federal de Educa\c c\~ao Tecnol\'ogica de Campos,
Campos dos Goytacazes, RJ 28030-130,  Brazil}

\begin{abstract}  
A static string in an AdS Schwarzschild space is dual to a heavy quark anti-quark pair
in a gauge theory at high temperature. This space is non confining in the sense that the energy is
finite for infinite quark anti-quark separation.
We introduce an infrared cut off in this space and calculate the corresponding string energy.
We find a deconfining phase transition at a critical temperature $\,T_C\,$. Above $\,T_C\,$ the string tension vanishes representing the deconfined phase. 
Below $\,T_C\,$ we find a linear confining behavior for large quark anti-quark separation.
This simple phenomenological model leads to the appropriate zero temperature limit, corresponding  to the Cornell potential and also describes a thermal deconfining phase transition. 
However the temperature corrections to the string tension do not recover the expected results for low temperatures. 
 
\end{abstract}


\maketitle

\vfill\eject
\section{Introduction}
Strong interactions at high energy are well described by QCD: Yang Mills SU(3)gauge theory
plus fermionic matter fields. However at low energies the theory is non perturbative because of the 
large coupling constant. 
Fundamental properties of strong interactions in this non perturbative regime, like confinement 
and mass generation are still open problems for theoretical physicists.

There are presently many indications that one can learn about strong interactions from gauge/string dualities. 
Important results have been obtained using this idea. 
Considering an $AdS$ space with an infrared cut off, Polchinski and Strassler 
calculated the scattering amplitudes for glueballs at high energies and fixed angles\cite{Polchinski:2001tt}.
They found the same scaling behavior (hard scattering)  of observed hadronic processes
known to be reproduced by QCD a long time ago\cite{QCD1,BRO}. 
This result opened the possibility of reconciling string theory with a phenomenological 
high energy behavior of strong interactions. 

The idea of introducing an infrared cut off in AdS space has been also  
useful to get results of hadronic spectroscopy.
Taking an AdS slice with boundary conditions as a phenomenological model with gauge/string duality 
the spectra of glueballs \cite{Boschi-Filho:2002vd,Boschi-Filho:2002ta}, 
light baryons and mesons \cite{deTeramond:2005su} were also obtained. 
Many other results in the recent literature give support to the idea of 
modeling aspects of  strong interactions using approximate gauge string dualities,
as for example \cite{Polchinski:2002jw,GI,BT,AN,Brodsky:2003px,Erlich:2005qh,Boschi-Filho:2005yh}.

An important tool for studying confinement in gauge theories is the Wilson loop.
Gauge string/duality can be used to calculate Wilson loops
from string configurations.
In the case of the AdS/CFT correspondence\cite{Maldacena:1997re,Gubser:1998bc,Witten:1998qj}
there is an exact gauge/string duality and the Wilson loop of a heavy quark anti-quark pair in 
the superconformal large N gauge theory is calculated
from a dual static string living in the AdS bulk\cite{RY,MaldaPRL}. 
The string lies along a geodesic with endpoints on the boundary 
representing the quark and anti-quark positions.
Its energy is proportional to the world sheet area 
and is a function of the quark anti-quark distance 
for an observer on the four dimensional boundary where the gauge theory is defined. 
In refs. \cite{RY,MaldaPRL} it was shown that for the AdS$_5$ space  
the energy shows a purely Coulombian (non confining) behavior, consistent  
with a conformal gauge theory. 
A discussion of Wilson loops associated with quark anti-quark potential
for different spaces can be found in \cite{Kinar:1998vq}. Other important related results can be found for
example in \cite{Greensite:1998bp,Greensite:1999jw,Bigazzi:2004ze,Martucci:2005yg}.

Approximate gauge/string duality is also useful for calculating Wilson Loops.
Inspired in the results of hadronic spectroscopy mentioned above, we calculated the world sheet 
area of a static string\cite{Boschi-Filho:2005mw} in an AdS space with infrared cut off 
assumed to represent a phenomenological approximation for a space dual to a gauge theory 
with a mass scale.
We found a potential energy with Coulomb like behavior for short distances 
and linear confining behavior for large distances.
This energy can be identified with the Cornell potential for the strong interaction
of a heavy quark anti-quark pair.

Here we build up an extension of this phenomenological model 
including thermal effects. We consider static strings in an  AdS Schwarzschild space
with an infrared cut off. This simple model leads to a deconfining phase transition at a critical 
temperature $T_C$.   
Above this temperature the potential energy remains finite for infinite quark separation, so the quarks are deconfined. 
Below $T_C$ the potential energy has a linear confining behavior.
Note however that for low temperatures our result for the dependence of the string tension with the temperature does not agree with lattice calculations\cite{Kaczmarek:1999mm,Petreczky:2005bd}. The expected result is that the string tension decreases with $\,-T^2 \,$  while our simple model gives a correction of 
$\,-T^4\,$. 
This is due to the fact that we are considering a background 
dual to a gauge theory at high temperatures. If we had considered the thermal AdS space, which is appropriate for low temperatures, we would get no corrections to the string tension.

\section{Quark anti-quark potential at finite temperature}  

A gauge/string duality involving a gauge theory at finite temperature 
was proposed by Witten in \cite{Witten:1998zw} inspired in the work of
Hawking and Page\cite{Hawking:1982dh}.
In this approach, for high temperatures, the AdS space accommodates a  Schwarzschild black hole and the
horizon radius is proportional to the temperature. For low temperatures the dual space would be an AdS space with compactified time dimension, known as thermal AdS. 
The gauge string duality using AdS Schwarzschild space has recently been applied to obtain the viscosity of a quark gluon plasma\cite{Policastro:2001yc,Kovtun:2004de}. 

Here we introduce a hard cutoff in the AdS Schwarzschild black hole metric as a phenomenological model for a space dual to a theory with both mass scale and finite (high) temperature. 
We consider this model to be an approximation for the temperatures in the range of interest, which is around
the deconfining critical temperature.
We place the cut off brane at the same position as in the
zero temperature case of ref. \cite{Boschi-Filho:2005mw}: $r = R$.  The corresponding metric is

\begin{equation}
\label{metric}
ds^2 \,=\, \Big( {r^2\over R^2} \Big) ( -\, f(r) \,dt^2 \,+ \, d{\vec x}^2 \,) +  
\Big( {R^2\over r^2} \Big) \,\frac{1}{f(r) } \, dr^2 \,+\, R^2 d^2 \Omega_5\,,
\end{equation}

\noindent where $R \le r < \infty \,$, $  f(r)\,=\, 1\,-\, r_T^4 / r^4 \,\ $ and the horizon radius $r_T\,$ 
is related to the Hawking temperature by $r_T\,=\, \pi\, R^2 \,T\,$. At zero temperature this space becomes an  Anti-de Sitter (AdS) slice. 
The problem of static strings in a space with metric (\ref{metric}), without any  cut off,
was discussed in detail in\cite{Rey:1998bq,Brandhuber:1998bs}. 
  
We will calculate the potential energy associated with a static string 
in the space (\ref{metric})  with endpoints at the boundary $r \to \infty\, $
separated by a coordinate distance $\Delta x\,=\, L$. 
The corresponding energy is proportional to the
world sheet area, calculated from the  Nambu-Goto action  
\begin{equation}
S \,=\,{1\over 2\pi \alpha^\prime}\, \int d\sigma d\tau 
\sqrt{ det \big( g_{MN} \partial_\alpha X^M 
\partial_\beta X^N \Big)\,\,}\,
\end{equation}

\noindent where $g_{MN}$ is the five dimensional sector of metric (\ref{metric}), excluding the
$S^5$ part. Solving the classical equations of motion, that corresponds to finding
the configuration that generates the world sheet with minimum area, one finds the energy of the quark anti-quark pair. 

Note that depending on the horizon radius (temperature) relative to the brane position $R$
we find different behaviors for the energy. 
If $r_T \ge R$ the  string will not be affected by the presence of the brane since it does not cross
the horizon. So the energy will be that described in ref. \cite{Rey:1998bq,Brandhuber:1998bs}
and there will be no confinement. If $r_T <  R$ the energy for large $L$ will grow linearly with
$L$ with a temperature dependent coefficient and the quarks are confined. The critical temperature 
corresponds to $r_T \,=\, R$.

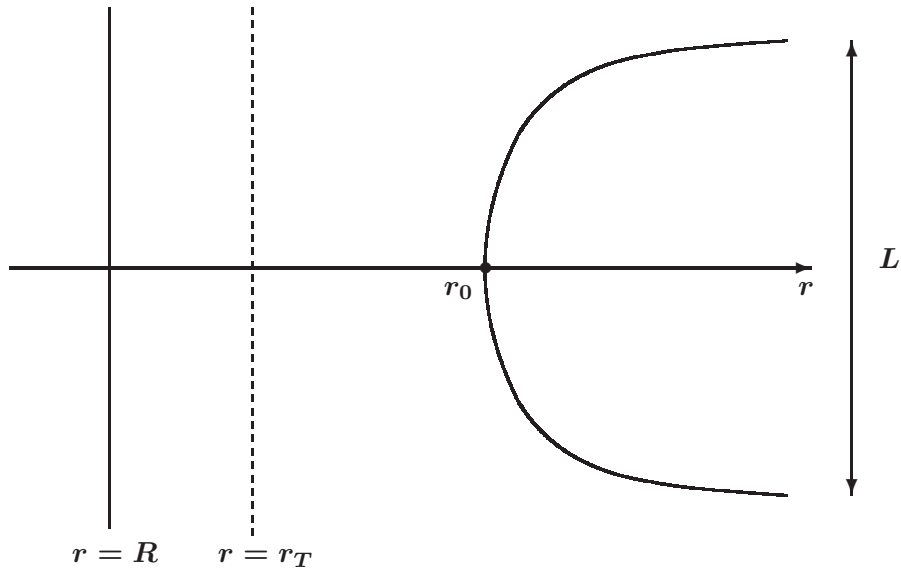
\begin{figure}
\centering

\
\setlength{\unitlength}{0.07in}
\vskip 5.5cm
{\begin{picture}(0,0)(18,0)
\rm\thicklines\bf
\put(50,-7){\vector(0,0){34}}
\put(50,27){\vector(0,-1){34}}
\put(52,10){\boldmath $ L $}
\put(-13,10){\vector(1,0){60}}
\put(46,8){\boldmath $ r $}
\put(19.5,8){\boldmath $ r_0 $}
\put(22.1,9.5){$\bullet$}
\bezier{600}(25,20)(20,10)(25,0)
\bezier{600}(25,0)(28,-5)(35,-6)
\bezier{600}(35,-6)(37,-6.5)(45,-7)
\bezier{600}(25,20)(28,25)(35,26)
\bezier{600}(35,26)(37,26.5)(45,27)
\multiput(5.2,-10)(0,1){40}{\line(0,1){0.5}}
\put(2.6,-12){\boldmath $ r=r_T $}
\multiput(-5.5,-9.5)(0,1){39}{\line(0,1){1}}
\put(-8.3,-12){\boldmath $ r=R $}
\end{picture}}
\vskip 2.5cm
\vskip .5cm
 \parbox{5in}{\caption{Schematic representation of a U-shaped geodesic with minimum $r_0$ far from the brane and from the horizon. This kind of geodesic appears for small quark anti-quark separation.}}
\end{figure}

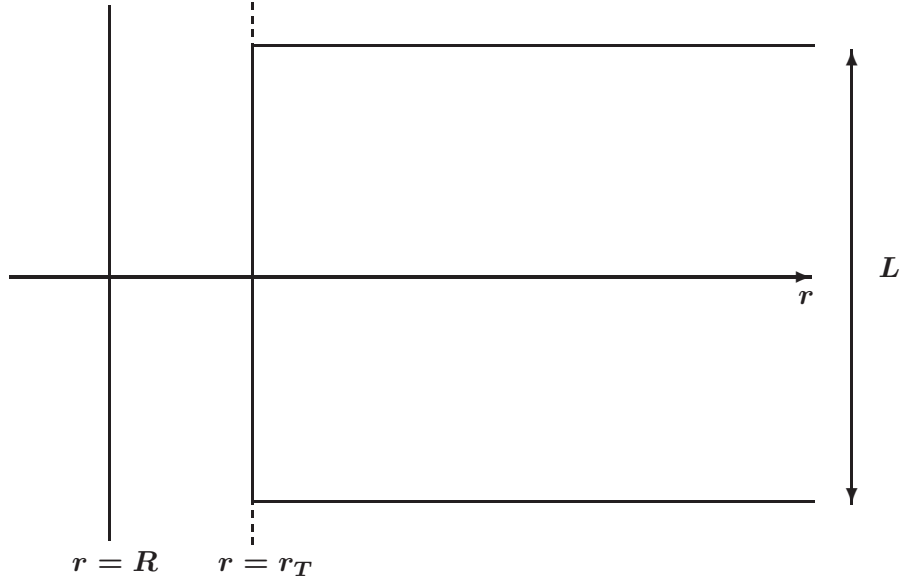
\begin{figure}
\centering

\
\setlength{\unitlength}{0.07in}
\vskip 5.5cm
{\begin{picture}(0,0)(18,0)
\rm\thicklines\bf
\put(50,-7){\vector(0,0){34}}
\put(50,27){\vector(0,-1){34}}
\put(52,10){\boldmath $ L $}
\put(-13,10){\vector(1,0){60}}
\put(46,8){\boldmath $ r $}
\put(5.2,-6.8){\line(1,0){42}}
\put(5.2,27.3){\line(1,0){42}}
\put(5.2,-6.7){\line(0,1){33.7}}
\multiput(5.2,-10)(0,1){41}{\line(0,1){0.5}}
\put(2.6,-12){\boldmath $ r=r_T $}
\multiput(-5.5,-9.7)(0,1){40}{\line(0,1){1}}
\put(-8.3,-12){\boldmath $ r=R $}
\end{picture}}
\vskip 2.5cm
\vskip .5cm
 \parbox{5in}{\caption{Schematic representation of a Box-shaped geodesic which reaches the horizon but not the brane. This is a typical situation at high temperature ($r_T>R$) and large quark anti-quark separation $L$.}}
\end{figure}

\begin{figure}
\centering

\
\setlength{\unitlength}{0.07in}
\vskip 5.5cm
{\begin{picture}(0,0)(18,0)
\rm\thicklines\bf
\put(50,-7){\vector(0,0){34.5}}
\put(50,27){\vector(0,-1){34.5}}
\put(52,10){\boldmath $ L $}
\put(-13,10){\vector(1,0){60}}
\put(46,8){\boldmath $ r $}
\bezier{600}(25.1,20)(25.1,10)(25.1,0)
\bezier{600}(25,0)(25.1,-5)(35,-6)
\bezier{600}(35,-6)(37,-6.5)(45,-7)
\bezier{600}(25,20)(25.1,25)(35,26)
\bezier{600}(35,26)(36,26.5)(45,27)
\multiput(4.2,-10)(0,1){40}{\line(0,1){0.5}}
\put(1.6,-12){\boldmath $ r=r_T $}
\multiput(25,-9.5)(0,1){39}{\line(0,1){1}}
\put(22.3,-12){\boldmath $ r=R $}
\end{picture}}
\vskip 2.5cm
\vskip .5cm
 \parbox{5in}{\caption{Schematic representation of a degenerated U-shaped geodesic with minima along the brane.}}
\end{figure}
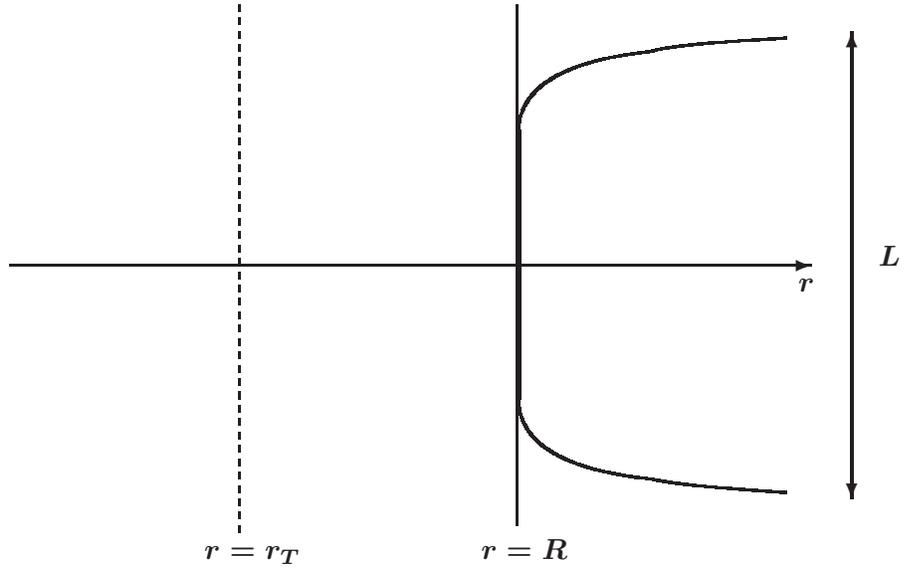

\subsection{ Deconfined phase: $r_T \ge R$}

For small quark anti-quark  separations $L$  the energy increases with $L$ 
up to a certain value  $L\,=\, L^\ast\,$. In this case the 
the world sheet with minimum area corresponds to a U-shaped  curve illustrated in figure {\bf  1}
with just one minimum value of coordinate $r = r_0  $, similar to those found 
in \cite{RY,MaldaPRL,Kinar:1998vq}. The energy in this case is 

\begin{equation}
\label{E1}
E \,=\,\frac{1}{ \pi \alpha^\prime} \Big\{ 
\int_1^{\infty }\,\Big( \frac{ \sqrt{y^4 - \frac{ r_T^4}{r_0^4 }}  }{\sqrt{ y^4 -1}}\,- 1 \Big)\,
r_0 dy - r_0  \Big\}\,,
\end{equation}

\noindent where we are subtracting the two quark masses, chosen as 
$ m_q\,=\,(1/2\pi\alpha^\prime ) \int_0^\infty dr\,$. The quark distance $L$ is related to $r_0$ by

\begin{equation}
\label{L}
L (r_0 ) \,=\,2\,\frac{R^2}{r_0} \sqrt{1- \frac{ r_T^4}{r_0^4 }}\, \int_1^{\infty}\,
\frac{ d y }{\sqrt{( y^4 -1)\,(y^4 - \frac{ r_T^4}{r_0^4 }) }}\,.
\end{equation}

\noindent From this equation one finds that $ L^\ast\,= \, 0.75 R^2/r_T\,$ since the U-shaped geodesics are energetically favorable for $r_0 \ge r_T/0.65 \,$. 

For quark distances $L$ larger than $ L^\ast$ the worlds sheet with minimum area corresponds to straight strings going directly to the horizon as the Box-shaped geodesic illustrated in figure {\bf 2}. 
In this situation the world sheet area becomes independent of $L$ since world sheets with spatial directions along the horizon have zero area. So the energy in this case is

\begin{equation}
E \,=\,-\, \frac{  r_T}{\pi \,\alpha^\prime } 
\end{equation}

\noindent according to our definition of quark masses.

\subsection{ Confined phase: $r_T <   R$}

If $r_T <  R$ the  situation changes since the string can reach the brane but not the horizon.
Actually, the horizon is not visible in the space in this case since we are considering 
the slice $r \ge R > r_T$.

For quark distances $L < L (r_0 )\vert_{r_0 = R} \,$ 
the string does not reach the brane and the energy is again 
given by (\ref{E1}). For quark distances larger than $ L (r_0 )\vert_{r_0 = R}\,$ the string touches the brane
and the geodesic will be a  U-shaped curve with degenerated minima.
That means two U-shaped halves connected by a straight line on the brane. These geodesics are illustrated 
in figure {\bf 3}. The energy is then  

\begin{equation}
\label{E2}
E \,=\,\frac{1}{ \pi \alpha^\prime} \left\{ 
\int_1^{\infty }\,\Big( \frac{ \sqrt{y^4 - \frac{ r_T^4}{R^4 }}  }{\sqrt{ y^4 -1}}\,- 1 \Big)\,
R dy - R  \,+ \frac{1}{2}\, \Big( L - L ( R) \Big) \, \sqrt{ 1 - \frac{ r_T^4 }{R^4}}   \right\}\,.
\end{equation}

\noindent For large values of the quark anti-quark separation $L$ the leading asymptotic 
behavior of the energy is linear: 

\begin{equation}
\label{linear}
 E \,\sim   \frac{ L}{2\pi \alpha^\prime} \,\, \sqrt{ 1 - \frac{ r_T^4 }{R^4}}\,\hskip3cm ( r_T < R).
\end{equation}

\noindent So there is confinement.    

As a remark, let us mention that for temperatures such that $  0.85 R <  r_T <  R$ there are three kinds of geodesics, depending on the distance $L$. 
For small $L$ the geodesics are the U-shaped curves given by eq. (\ref{L}). Note that
the expression (\ref{L}) leads to values of  $ L$ in a finite range: $ 0 \le L \le L_{Max} ( r_T)  $.
For very large values of $L$  the geodesics are the box-shaped straight lines going to the brane and 
on the brane. So there is confinement in this regime. 
For intermediate values of $L$ the geodesic will be the U-shaped curve with degenerated minima
with a straight line parallel to the brane but not at $r = R$. Note that
when the distance $L$ reaches some large value the path with straight lines becomes the geodesic.
So, the asymptotic behavior for $L \to \infty $ in this temperature range will be confining with again a linear term given by  eq. (\ref{linear}).

\section{ Conclusions}

We show in figure {\bf 4} the energies obtained, using the expressions presented in the previous section, 
for temperatures $ T =0, \, \,T = 0.8 T_C \,\,, T = T_C\,\,$ and $\,\, T = 2 T_C\,$. This figure
is similar to that found from lattice in \cite{Petreczky:2005bd}.
This illustrates the fact that in our model the energy of static strings associated with the quark anti-quark potential present a confining behavior for 
temperatures below  $ T_C \,=\, 1 /\, \pi\, R\,$. In this case there is a linear term in the energy, for large quark distances $L\,\,$ given by  $\,E \sim  \sigma (T)\,L\,$  with 

\begin{equation}
\label{tensionT}
\sigma (T) \,=\,\frac{1}{2 \pi \alpha^\prime} \sqrt{ 1 -  (\pi R T )^4 }\,\,\hskip3cm ( T < T_C ).
\end{equation}
 
\noindent At zero temperature this coefficient is identified with the string tension of the Cornell potential 
$ 1/ ( 2 \pi \alpha^\prime ) \,=\, 0.182$ Gev$^2$ and  
taking the AdS radius to have the value $ R = 1.4 \,{ \rm GeV}^{-1}\,$ 
\cite{Boschi-Filho:2005mw}, we find a critical temperature $ T_C \,\sim \,{\rm  230 MeV}\,$.

If we extrapolate our model to low temperatures it would imply  corrections of order $\,- \,T^4 \,\, $ to the string tension, as can be seen from  eq. (\ref{tensionT}). 
However\footnote{We thank Michael Teper for calling our attention to this point.} the results from fluctuations of strings in flat space at finite temperature\cite{Pisarski:1982cn}
and from lattice calculations\cite{deForcrand:1984cz,Kaczmarek:1999mm} imply corrections  
of order $\,-\, T^2\,$ at low temperatures. This shows that our simple model is not appropriate for low temperatures.

There is a very recent result by Herzog\cite{Herzog:2006ra} which indicates that the dual space 
for temperatures below $T_C$ should be indeed the thermal AdS. 
If we had considered this metric for low temperatures, instead of the AdS Schwarzschild black hole, we would get no thermal corrections to the string tension. 
So the confined phase would have the same string tension as in the zero temperature case.
It would highly desirable to find a holographic phenomenological model that gives the expected low temperature corrections.
 
It is interesting to mention the recent article \cite{Ghoroku:2005kg} that uses the cut off AdS Schwarzschild space to describe light mesons at finite temperature. Another recent article \cite{Andreev:2006eh} uses a deformed AdS Schwarzschild space to obtain the spatial string tension at finite temperature.

\noindent 

\begin{figure}
\centering
\includegraphics[width=8cm]{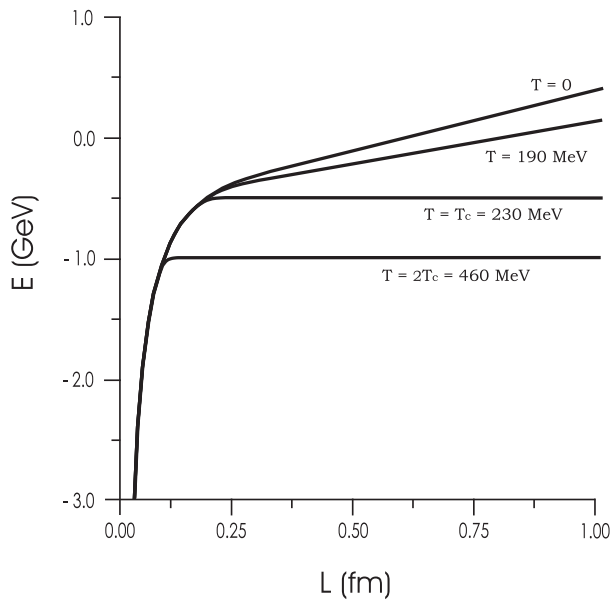}
\parbox{5in}{\caption{ Energy as a function of string end-points separation for different temperatures. } }
\end{figure}

\vskip 1cm

\noindent {\bf Acknowledgments}: We would like to thank Eduardo Fraga, 
Silvio Sorella and the participants of the workshop ``Infrared QCD in Rio",  where the preliminary results of this article were presented, for very interesting comments.  We would like to thank also Michael Teper,  Andrei Starinets and the other participants of the ``Hadrons and Strings" Trento workshop for very interesting discussions.
The authors are partially supported by CNPq and Faperj.

\end{document}